\documentclass{article}

\usepackage{PRIMEarxiv}

\usepackage[utf8]{inputenc} 
\usepackage[T1]{fontenc}    
\usepackage{hyperref}       
\usepackage{url}            
\usepackage{multirow,booktabs}       
\usepackage{amsfonts}       
\usepackage{nicefrac}       
\usepackage{microtype}      
\usepackage{lipsum}
\usepackage{fancyhdr}       
\usepackage{graphicx}       
\usepackage{comment}       
\graphicspath{{media/}}     

\usepackage{amsmath}

\title{Phenotyping Tibial Plateau Fractures via Self-Supervised Learning: A Label-Agnostic Framework with Expert Validation
}

\author{
  Miral Elnakib \\
  Faculty of Sciences \\
  Alexandria University \\
  \texttt{miralelnakib7@gmail.com} \\
  \And
  Muhammad Saad \\
  Faculty of Sciences \\
  Alexandria University \\
  \texttt{m.saad@alexu.edu.eg} \\
  \And
  Ahmad Al-Kabbany \\
  Multimedia Interaction and Communication Lab \\
  Wearables, Biosensing, and Biosignal Processing Research Lab \\
  Arab Academy for Science and Technology \\
  \texttt{alkabbany@ieee.org, alkabbany@aast.edu} \\
}

\begin{document}
\maketitle

\begin{abstract}
The full potential of artificial intelligence in tibial plateau fracture 
characterisation remains unrealised, constrained by a fundamental 
dependency on labelled datasets whose consistency cannot be guaranteed: 
conventional classification schemes such as Schatzker and AO/OTA suffer 
from well-documented inter-observer variability, causing supervised 
models to learn human disagreement rather than stable fracture 
morphology. We design, implement, and validate a label-agnostic framework 
that eliminates this constraint by learning fracture representations 
directly from imaging data without observer-assigned labels. A 
RadImageNet-pretrained ResNet-50 encoder is fine-tuned on $154$ cleaned 
knee radiographs using the SimCLR contrastive objective, preceded by a 
structured eight-step data cleaning protocol to remove acquisition-driven 
confounds, and followed by UMAP dimensionality reduction and $k$-means 
clustering to discover four candidate imaging-derived phenotypes. 
Phenotype validity is assessed through a blinded expert review protocol 
administered independently to two 
clinicians\footnote{Prof.\ Ahmed Fouad Shams El-Din Mostafa, 
Professor of Orthopaedic Surgery at Menoufia University, Egypt, and Dr.\ Islam Mohamed Fouad El-Gohary, Consultant Orthopedic Surgeon, Egypt}, providing qualitative validation decoupled from 
the quantitative pipeline. The four phenotypes demonstrate robust 
stability (bootstrap $\text{ARI} = 0.319 \pm 0.041$), strong internal 
cohesion (silhouette $= 0.511$), and coherence ratings ranging from 
$3$ to $5$ out of $5$ from both expert reviewers under blinded 
conditions, with all four groups meeting the pre-specified face validity 
threshold; one phenotype was unanimously identified by both reviewers 
as exhibiting comminution --- a high-complexity feature the pipeline 
isolated without any supervisory signal. Inter-partition comparison 
against Schatzker labels yields $\text{ARI} = 0.013$, confirming 
orthogonality to conventional classification boundaries while expert 
responses confirm genuine morphological content. Notably, expert reviewers anchored to established classification 
vocabularies perceived imaging-derived groups as heterogeneous 
precisely where inter-partition alignment with Schatzker was lowest, 
suggesting that Schatzker-trained perception and label-agnostic 
embedding geometry are measuring orthogonal dimensions of the same 
morphological space. These findings establish 
label-agnostic SSL phenotyping as a reproducible and clinically 
interpretable complement to conventional fracture classification, with 
implications for treatment standardisation and data-driven orthopaedic 
decision support.
\end{abstract}

\keywords{tibial plateau fractures \and self-supervised learning \and 
unsupervised phenotyping \and contrastive learning \and fracture 
morphology \and RadImageNet \and clustering \and label-agnostic learning 
\and musculoskeletal imaging \and inter-observer variability}

\section{Introduction}
\label{sec:intro}

Artificial intelligence is reshaping diagnostic medicine across 
specialties~\cite{rajpurkar2022ai}, and orthopaedic surgery is no 
exception: deep learning models have demonstrated clinically meaningful 
performance on fracture detection, implant planning, and post-operative 
assessment from plain radiographs and cross-sectional 
imaging~\cite{olczak2017ai, cheng2022ai}.

Among orthopaedic injuries, tibial plateau fractures (TPFs) occupy a 
position of particular clinical importance. Affecting the proximal 
articular surface of the tibia, these injuries involve a complex 
interplay of split and depression components, articular comminution, 
metaphyseal extension, and bicondylar involvement, and their management 
--- ranging from conservative treatment to complex multi-implant 
fixation --- depends critically on an accurate characterisation of 
fracture morphology. In routine clinical practice and research, this 
characterisation is mediated through established classification schemes. 
The Schatzker system~\cite{schatzker1979} and the AO/OTA 
classification~\cite{marsh2007aota} are by far the most widely adopted, 
providing a shared vocabulary for treatment planning, surgical 
communication, and research stratification. Additional systems have been 
proposed, including the Hohl and Moore 
classification~\cite{hohlmoore1990} and, more recently, the three-column classification of Luo et al.~\cite{luo2010} for complex bicondylar injuries, reflecting ongoing efforts to capture fracture 
complexity more precisely.

Despite their clinical utility, conventional classification schemes 
share a fundamental limitation that constrains their use as 
supervisory signals for machine learning: they depend on observer 
judgement applied to inherently ambiguous radiographic appearances, 
and inter-observer agreement is known to be 
imperfect~\cite{chabok2003,zhu2013}. Different clinicians 
examining the same radiographic study may assign different Schatzker 
classes, particularly for borderline or complex cases at the boundaries 
between categories. This inter-observer variability introduces a 
critical methodological tension for supervised learning: a model trained 
to reproduce observer-assigned labels does not learn stable fracture 
morphology --- it learns to replicate the disagreement embedded in those 
labels. Performance metrics computed against inconsistent labels may 
consequently underrepresent true morphological understanding and fail 
to translate to improved clinical decision support. The problem is not 
merely technical: it reflects a deeper epistemological constraint on 
label-dependent learning when the labels themselves are noisy by 
construction. What is needed is an approach that learns directly from 
imaging data, without relying on the consistency of observer-assigned 
categories --- a problem that is, by construction, one of 
self-supervised representation learning and unsupervised clustering.

Self-supervised learning (SSL) has emerged as a powerful paradigm for 
learning visual representations without labels, demonstrating strong 
performance across natural and medical imaging 
domains~\cite{chen2020simclr, caron2021dino, azizi2021simclr}. In 
medical imaging, SSL methods have been applied to chest 
radiographs~\cite{azizi2021simclr}, dermatology~\cite{azizi2021simclr}, 
and pathology slides~\cite{steyaert2023multimodal}, with domain-adaptive 
pretraining strategies such as RadImageNet~\cite{mei2022radimagenet} 
showing particular promise for small-dataset musculoskeletal tasks. 
Unsupervised clustering for clinical phenotype discovery has likewise 
attracted growing interest as a means of identifying data-driven patient 
subgroups that complement or refine conventional 
classifications~\cite{steyaert2023multimodal}. However, the specific 
application of label-agnostic SSL phenotyping to traumatic fracture 
morphology --- and to tibial plateau fractures in particular --- remains 
unexplored.

\begin{figure*}[t]
    \centering
    \includegraphics[width=\textwidth]{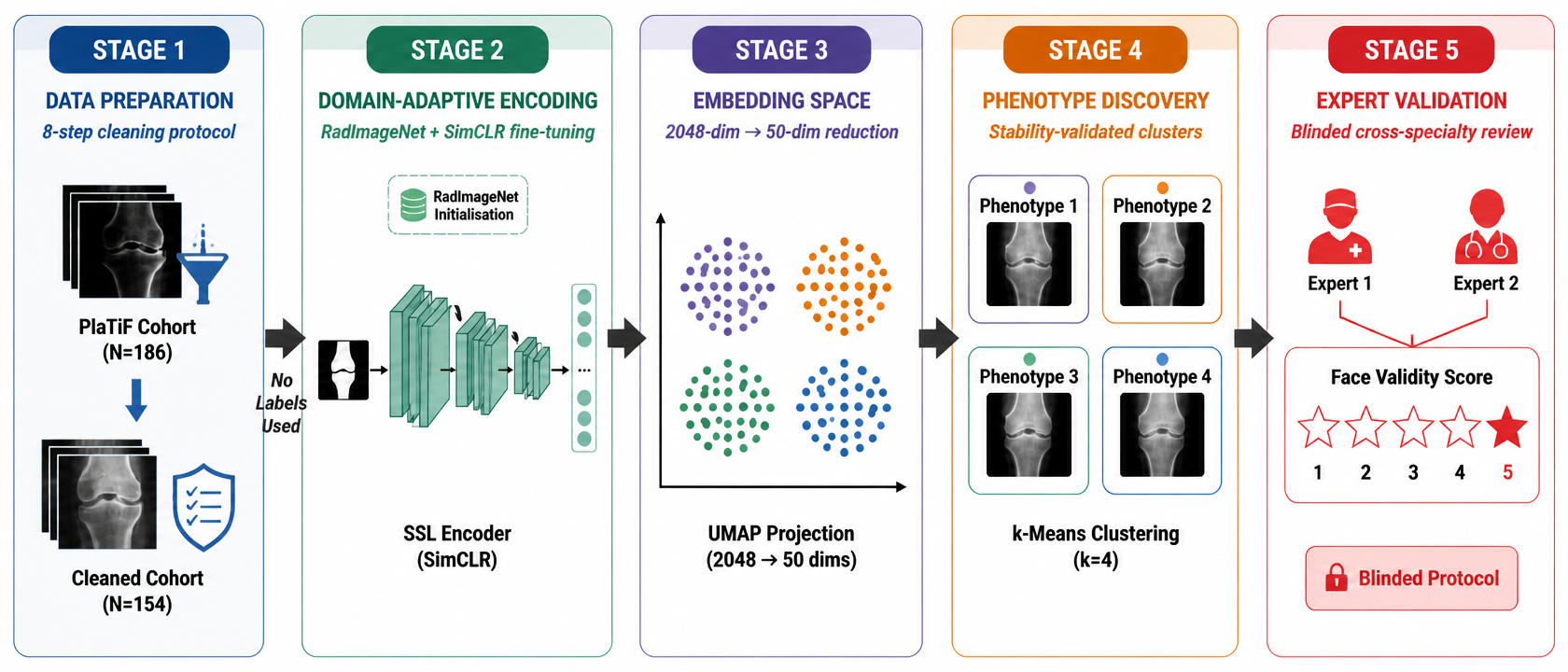}
    \caption{Overview of the proposed label-agnostic SSL phenotyping 
  pipeline for tibial plateau fractures. The pipeline proceeds through 
  five stages: data preparation and cleaning, domain-adaptive 
  self-supervised representation learning, UMAP dimensionality reduction, 
  unsupervised phenotype discovery via k-means clustering, and blinded 
  clinical face validity assessment by independent expert reviewers.}
    \label{fig:pipeline}
\end{figure*}

This work addresses that gap. We present a label-agnostic imaging 
pipeline that learns fracture representations directly from knee 
radiographs using a RadImageNet-pretrained ResNet-50 encoder trained 
with the SimCLR contrastive SSL objective, without any reference to 
Schatzker or AO/OTA labels during training. The learned embeddings are 
subsequently clustered to discover candidate imaging-derived phenotypes, 
which are evaluated for geometric stability, reproducibility, and clinical face validity through a blinded expert validation protocol involving two independent expert reviewers with relevant clinical experience. The comparison 
against conventional labels is performed entirely post-hoc, as an 
inter-partition alignment analysis rather than an accuracy evaluation, 
preserving the label-agnostic integrity of the pipeline throughout. We depict the proposed TPF phenotyping pipeline in Fig.~\ref{fig:pipeline}. The contributions of this work are as follows:

\begin{enumerate}

    \item We propose and validate a complete label-agnostic SSL 
    pipeline for tibial plateau fracture phenotyping from plain 
    radiographs, providing the first demonstration of imaging-derived 
    fracture phenotype discovery for this injury type without reliance 
    on observer-assigned classification labels.

    \item We introduce a domain-adaptive encoder initialisation strategy 
    for small-dataset orthopaedic imaging, demonstrating that 
    RadImageNet-pretrained convolutional weights provide a more 
    appropriate initialisation than ImageNet pretraining for SSL 
    fine-tuning at dataset sizes typical of rare traumatic injury 
    cohorts, and providing a principled justification for architecture 
    selection under data scarcity constraints.

    \item We present a structured eight-step data cleaning protocol for 
    musculoskeletal radiograph cohorts in unsupervised learning 
    pipelines, with documented exclusion criteria, versioned splits, 
    and a reproducibility checklist, addressing a methodological gap 
    in the clinical AI literature where dataset provenance is rarely 
    reported with sufficient transparency.

    \item We demonstrate that imaging-derived phenotypes are 
    geometrically stable (bootstrap $\text{ARI} = 0.319 \pm 0.041$), 
    internally coherent (KMeans silhouette $= 0.511$), and clinically 
    interpretable, receiving coherence ratings of $3$--$5$ out of $5$ 
    from two independent expert reviewers under blinded conditions, with 
    all four phenotypes meeting the pre-specified face validity threshold 
    and unanimous cross-reviewer agreement on comminution as an 
    exclusively high-complexity descriptor confined to a single 
    imaging-derived phenotype.

    \item We show that imaging-derived phenotypes are orthogonal to 
    conventional Schatzker classification boundaries 
    ($\text{ARI} = 0.013$, $\text{NMI} = 0.053$), while simultaneously 
    capturing clinically recognisable morphological structure that 
    expert reviewers spontaneously characterised using Schatzker-type 
    descriptors --- establishing that the two organisational frameworks 
    are complementary rather than competitive, and that the near-zero 
    inter-partition alignment reflects a difference in organisational 
    principle rather than an absence of clinical signal.

\end{enumerate}

\section{Related Work}
\label{sec:related}

\subsection{Supervised Deep Learning for Tibial Plateau Fracture
Classification}

The dominant paradigm in computational analysis of tibial plateau
fractures has been fully supervised deep learning, in which
convolutional neural networks are trained to reproduce
observer-assigned Schatzker or AO/OTA class labels from radiographic
or CT inputs. The results of this line of work are instructive
precisely because they reveal the ceiling imposed by the label
consistency problem that motivates our approach.

Van der Gaast et al.~\cite{vandergaast2025} trained GoogleNet and
ResNet architectures on 1,506 knee radiographs from 753 patients
across multiple centres for simultaneous fracture detection and
Schatzker classification. While fracture detection achieved high
sensitivity (92.7\%), Schatzker classification accuracy reached only
34.6\%, leading the authors to explicitly conclude that supervised
models may benefit from abandoning the Schatzker system in favour of
alternative organisational frameworks, citing the scheme's low
inter-observer agreement on plain radiographs as a fundamental
limiting factor. Huo et al.~\cite{huo2025} developed a
MobileNetV3--YOLOv8 model trained on 3,547 radiographs from five
centres for TPF detection, including occult fractures, demonstrating
that AI assistance can improve sensitivity and reduce interpretation
time for less experienced physicians --- yet their work, like most in
this space, relies entirely on expert-annotated ground truth whose
consistency is assumed rather than quantified. Cai
et al.~\cite{cai2024unet} approached the problem from CT rather than
radiograph, training a 3D U-Net on 234 cases to automatically segment
the tibial plateau and generate fracture maps that assist residents
with Schatzker classification, reporting Dice coefficients exceeding
0.950 and measurable improvements in resident classification accuracy
and efficiency. Together, these studies establish that supervised
deep learning can reliably detect and segment tibial plateau
fractures, while consistently falling short on Schatzker
classification --- a pattern that is not attributable to architectural
limitations but to the intrinsic noise in the supervisory signal
itself. Our work takes this observation as its starting point rather
than its conclusion, proposing a pipeline that circumvents the
labelling bottleneck by design rather than attempting to engineer
around it.

\subsection{Reducing Annotation Dependency in Fracture Imaging}

A parallel line of work has sought to reduce the annotation burden
in fracture analysis by leveraging self-supervised pretraining or
semi-supervised learning strategies that make use of unlabeled or
partially labelled data. This direction shares the motivating
observation that expert annotation is scarce, expensive, and
variable, but addresses it through a different architectural response:
using labels sparsely rather than eliminating them entirely.

Yue et al.~\cite{yue2025mae} proposed an MAE-based pretraining
strategy for TPF segmentation in CT, leveraging masked image
modelling on unlabeled CT volumes to capture global skeletal
structure and fine-grained fracture detail before fine-tuning on a
small set of annotated cases. Their method achieved a Dice
coefficient of 95.81\% with only 20 annotated training cases,
substantially outperforming conventional semi-supervised baselines
and demonstrating strong transferability to an independent hip
fracture dataset. This work is methodologically adjacent to ours in
its use of self-supervised pretraining to extract fracture-relevant
representations from unlabeled data, and in its explicit motivation
by annotation scarcity. The distinction is fundamental, however:
Yue et al.\ treat label reduction as a practical engineering goal
--- minimising the number of annotations required to reach a
segmentation performance target --- while retaining the supervised
fine-tuning step as an essential component. Our pipeline eliminates
label dependency entirely from both the representation learning and
the cluster discovery stages, positioning the absence of labels not
as a practical convenience but as a methodological prerequisite for
avoiding the consistency ceiling that supervised objectives inherit
from their training labels. \textit{Where Yue et al.\ ask how few labels are
needed, we ask what structure can be discovered without any}.

Taken together, these two directions --- supervised classification
bounded by label noise, and annotation-efficient learning that
reduces but does not eliminate label dependency --- define the
landscape into which the present work intervenes. \textbf{To the best of our knowledge, no prior work has applied a fully label-agnostic SSL
phenotyping framework to tibial plateau fractures, treating the
discovered cluster structure as the primary scientific contribution rather than as a preprocessing step toward a downstream supervised objective}.

\section{Methodology}
\label{sec:methodology}

\subsection{Problem Formulation}
\label{subsec:problem}

Tibial plateau fracture (TPF) classification in routine clinical practice relies on 
observer-assigned categorical labels drawn from established schemes, principally the 
Schatzker system and the AO/OTA classification. These schemes partition fracture 
morphology into discrete classes based on qualitative radiographic features, and are 
widely used for treatment planning, surgical communication, and research stratification. 
However, both systems are subject to well-documented inter-observer variability, 
particularly in borderline and complex cases, where the same radiographic study may 
receive different class assignments from different clinicians. This inconsistency 
introduces a fundamental methodological tension for supervised machine learning: a model 
trained to reproduce observer-assigned labels learns to replicate the human disagreement 
embedded in those labels rather than the stable underlying morphological structure the 
labels were intended to represent.

This work adopts a label-agnostic formulation that sidesteps this tension entirely. 
Rather than framing the problem as classification --- mapping an image to a predefined 
class --- we frame it as representation learning followed by unsupervised phenotype 
discovery. Formally, let $\mathcal{X} = \{x_i\}_{i=1}^{N}$ denote a cohort of $N$ 
tibial plateau fracture radiographs, where no label information is used during model 
training. The objective is to learn an encoder
\begin{equation}
    f_{\theta} : \mathcal{X} \rightarrow \mathbb{R}^{d},
    \label{eq:encoder}
\end{equation}
that maps each image $x_i$ to a compact embedding vector $z_i \in \mathbb{R}^{d}$ such 
that geometrically similar embeddings correspond to radiographs sharing meaningful 
morphological structure. The encoder parameters $\theta$ are optimised entirely from 
the imaging data via a self-supervised learning objective, without any reference to 
Schatzker or AO/OTA labels.

Once a trained encoder is obtained, the embedding matrix 
$Z = \{z_i\}_{i=1}^{N} \subset \mathbb{R}^{d}$ is subjected to unsupervised clustering 
to partition the cohort into $k$ candidate phenotypes 
$\mathcal{C} = \{C_1, C_2, \ldots, C_k\}$, where
\begin{equation}
    \bigcup_{j=1}^{k} C_j = \mathcal{X}, 
    \qquad 
    C_j \cap C_l = \emptyset \quad \forall\, j \neq l.
    \label{eq:partition}
\end{equation}
The optimal $k$ is determined empirically from internal cluster quality metrics rather 
than prescribed by any existing classification scheme. Each resulting phenotype is 
subsequently characterised in terms of its morphological content and evaluated for 
stability, internal geometric coherence, and alignment with conventional labels.

Crucially, the comparison against Schatzker and AO/OTA labels is performed entirely 
post-hoc, as an external characterisation step, and the labels play no role in shaping 
the learned representations or the cluster boundaries. Schatzker class assignments are 
treated not as ground truth but as an independent reference partition reflecting 
conventional clinical categorisation, against which the degree of alignment --- rather 
than classification accuracy --- is quantified. This design ensures that any structure 
discovered by the pipeline reflects genuine imaging-derived morphological organisation 
rather than a reproduction of observer-assigned categorical boundaries.

\subsection{Dataset}
\label{subsec:dataset}

\subsubsection{Source and Composition}
The primary dataset used in this study is the Plateau Tibial Fracture 
(PlaTiF) dataset~\cite{kazemi2026platif}, a publicly available annotated collection 
of knee radiographs assembled for tibial plateau fracture research. The 
dataset comprises 421 radiographs acquired from 186 patients, accompanied 
by Schatzker class annotations and patient-level metadata including age, 
mechanism of injury, and treatment type. Radiographs were acquired across 
multiple views, including anterior-posterior (AP), lateral, and oblique 
projections, and are provided in MATLAB \texttt{.mat} format with 
associated metadata in a structured spreadsheet. \textit{The PlaTiF dataset 
represents the only publicly available dedicated tibial plateau fracture 
radiograph collection at the time of this study, and was selected 
accordingly as the sole imaging source for this work}.

\subsubsection{Motivations and Limitations}
The decision to use PlaTiF as the primary data source was driven by 
several considerations. Its public availability ensures full 
reproducibility of the pipeline without dependence on proprietary 
institutional archives. The availability of Schatzker annotations, while 
not used during model training, enables the post-hoc inter-partition 
comparison described in Section~\ref{sec:eval}. Finally, the dataset's 
exclusive focus on tibial plateau injuries avoids the anatomical 
heterogeneity that would arise from repurposing general musculoskeletal 
radiograph collections for this specific fracture type.

These advantages must be weighed against several limitations that bear 
directly on the interpretation of results. The cohort of 186 patients is 
substantially smaller than the dataset sizes typically employed in 
self-supervised learning pipelines, where tens of thousands of images are 
conventional. This constraint motivated key architectural decisions 
described in Section~\ref{sec:ssl}, specifically the use of a 
domain-adapted convolutional encoder initialised from a large medical 
imaging pretraining corpus rather than a data-hungry transformer 
architecture. Furthermore, the dataset originates from a single 
institution, limiting the diversity of imaging acquisition parameters, 
scanner characteristics, and patient demographics represented in the 
training cohort. External validation on an independent multi-centre 
cohort, which is deferred to a subsequent phase of this work, will be 
necessary before generalisability claims can be made.

\subsubsection{Data Cleaning and Cohort Definition}
\label{sec:data_cleaning}
The raw PlaTiF cohort of 186 cases underwent a structured eight-step 
cleaning protocol prior to any model training, motivated by the 
observation that unsupervised clustering is particularly sensitive to 
acquisition-driven artefacts: without label supervision to override 
spurious signals, a contaminated training set risks producing clusters 
that reflect imaging conditions rather than fracture morphology.

The protocol proceeded as follows. First, a paginated visual gallery of 
all 186 cases was generated, displaying each image alongside its patient 
identifier and Schatzker class label, to enable systematic case-by-case 
review. Second, a trained reviewer conducted a manual assessment of 
anatomy and view type, flagging cases that did not show the knee or 
proximal tibia, or for which no anterior-posterior or near-AP frontal 
view was available. Third, a dedicated hardware review pass identified 
cases with visible post-operative implants, including intramedullary 
nails, locking plates, external fixators, and surgical wires. Fourth, an 
automated quality check flagged cases with extreme exposure values, 
near-zero image contrast, failed region-of-interest detection, or 
sub-threshold image resolution. Fifth, perceptual hashing was used to 
identify near-duplicate studies, with confirmed duplicates resolved by 
retaining the higher-quality instance. Sixth, all cases flagged as 
uncertain across the preceding steps were compiled into a secondary 
review gallery and adjudicated by the third author (Principal Investigator-PI) of this article, whose decisions were recorded 
with initials and date in a versioned exclusion log.

Applying this protocol to the 186-case cohort yielded 32 confirmed 
exclusions, distributed across the following categories: wrong anatomy or 
field of view ($n = 18$), post-operative hardware ($n = 8$), non-AP view 
only ($n = 3$), failed automated quality check ($n = 1$), and confirmed 
duplicate ($n = 1$). An additional $n = 1$ case was excluded following a 
post-protocol PI review of cluster montages that identified residual 
anatomy concerns not captured by the initial gallery review. The final 
clean cohort comprised $N = 154$ cases.

\subsubsection{Dataset Splits}
The 154-case cohort was partitioned at the patient level into training, 
validation, and test subsets using a fixed random seed to ensure 
reproducibility. The training set comprised 107 cases (70\%), the 
validation set 15 cases (10\%), and the test set 32 cases (20\%). 
Patient-level splitting was enforced throughout to prevent data leakage 
between subsets. Split assignments were saved in a versioned manifest 
file and held fixed for all experiments reported 
in this work. Self-supervised training was performed exclusively on the 
107-case training split; embeddings were subsequently extracted for all 
154 cases prior to clustering, ensuring that the full cohort contributed 
to phenotype discovery despite the label-free training objective.

\subsection{Preprocessing and Augmentation}
\label{subsec:preprocess}

\subsubsection{Preprocessing Pipeline}
Raw images were extracted from the PlaTiF \texttt{.mat} archive files 
and converted to single-channel greyscale arrays prior to any further 
processing. Each image underwent a sequence of standardisation steps 
designed to reduce acquisition-driven variability while preserving the 
morphological signal relevant to fracture characterisation.

Region-of-interest (ROI) cropping was applied to centre the tibial 
plateau within the image frame, using an Otsu-threshold-based detector 
to identify the bone-containing region followed by a fixed-margin 
expansion to retain periarticular context. All images were subsequently 
resized to a canonical resolution of $224 \times 224$ pixels to satisfy 
the input requirements of the ResNet-50 encoder. Orientation 
normalisation was applied to standardise left-right laterality across 
cases, ensuring that the encoder was not exposed to mirror-image 
variation that could introduce spurious asymmetry cues into the learned 
representations.

Intensity normalisation was performed at the population level rather 
than per image. Specifically, the mean $\mu_{\text{train}}$ and standard 
deviation $\sigma_{\text{train}}$ of pixel intensities were computed 
exclusively from the 107-case training split, and all images --- 
including validation and test cases --- were normalised according to
\begin{equation}
    \hat{x}_i = \frac{x_i - \mu_{\text{train}}}{\sigma_{\text{train}} 
    + \epsilon},
    \label{eq:normalization}
\end{equation}
where $\epsilon$ is a small constant included for numerical stability. 
This population-level normalisation strategy was deliberately chosen over 
per-image min-max normalisation, which would have destroyed the 
cross-patient variation in bone density --- a potentially informative 
morphological signal --- by mapping every image independently to the 
same intensity range. Computing normalisation statistics from the 
training split only ensures that validation and test cases are treated 
as unseen data throughout the pipeline, consistent with standard 
machine learning evaluation practice.

\subsubsection{Augmentation Strategy}
\label{sec:aug}

Data augmentation in self-supervised learning serves a dual role: it 
defines the invariances the encoder is trained to respect, and it 
effectively expands the training distribution seen during optimisation. 
In the context of fracture radiographs, however, augmentation design 
requires particular care. Aggressive photometric or geometric transforms 
that are standard in natural image SSL --- such as heavy blurring, 
extreme cropping, or large-scale colour distortion --- risk destroying 
the subtle visual cues that distinguish fracture morphology, including 
fine cortical disruption lines, articular step-off contours, and 
metaphyseal extension patterns. The augmentation policy adopted in this 
work was therefore deliberately conservative, prioritising preservation 
of fracture-relevant structure over maximal distributional diversity.

Geometric augmentations comprised small random rotations within 
$\pm 15^{\circ}$, random affine translations of up to $10\%$ of the 
image dimension in each axis, and symmetric reflect-mode padding of 
$22$ pixels prior to rotation and translation to avoid black-border 
artefacts at image boundaries. Photometric augmentations comprised 
colour jitter with brightness and contrast perturbation limited to 
$\pm 8\%$, and random Gaussian blur with kernel size $3 \times 3$ and 
sigma uniformly sampled from $[0.1, 0.5]$, applied with probability 
$0.3$. All augmented images were clipped to the range $[0.05, 0.95]$ 
after transformation to suppress near-black border cutoffs introduced 
by geometric transforms. A content validation check was applied to each 
augmented view, rejecting and resampling any crop in which more than 
$20\%$ of pixels fell below $0.05$, ensuring that augmented views 
retained sufficient bone signal for meaningful representation learning.

For the SimCLR objective described in Section~\ref{sec:ssl}, each 
training image $x_i$ was passed through the augmentation pipeline twice 
independently to produce a pair of views $(\tilde{x}_i^{(1)}, 
\tilde{x}_i^{(2)})$, constituting a positive pair. The encoder was 
trained to produce similar representations for positive pairs while 
differentiating them from representations of other images in the same 
batch, as formalised in Section~\ref{sec:ssl}. The augmentation 
parameters were held fixed across all training runs to ensure 
reproducibility.

\subsection{Self-Supervised Representation Learning}
\label{sec:ssl}

\subsubsection{Architecture Selection and Initialisation}
The choice of encoder architecture for self-supervised representation 
learning in small-dataset medical imaging settings involves a fundamental 
trade-off between model expressivity and the availability of appropriate 
pretraining initialisation. Vision Transformers (ViTs), which have 
demonstrated state-of-the-art representation quality in large-scale SSL 
settings~\cite{he2022mae, caron2021dino}, achieve their performance 
advantage by learning spatial relationships entirely from data, without 
the inductive biases of local connectivity and translation equivariance 
that are built into convolutional architectures. This absence of 
inductive bias is a liability rather than an asset when the fine-tuning 
dataset is small: with 107 training images, a ViT cannot acquire the 
spatial priors it requires to produce well-structured representations, 
and the resulting embeddings risk reflecting superficial statistics of 
the training distribution rather than meaningful morphological structure.

Convolutional architectures, by contrast, embed translation equivariance 
and local feature hierarchies directly into their inductive structure, 
making them substantially more data-efficient in low-data regimes. A 
ResNet-50 backbone~\cite{he2016resnet} was therefore selected as the 
encoder for this work. Critically, this architectural choice was 
enabled by the availability of domain-appropriate pretraining 
initialisation. The encoder was initialised from weights pretrained on 
RadImageNet~\cite{mei2022radimagenet}, a large-scale open-access 
radiological imaging dataset comprising 1.35 million annotated images 
spanning CT, MRI, and ultrasound modalities across musculoskeletal, 
neurological, oncological, and other pathological categories. 
RadImageNet-pretrained models have been shown to outperform 
ImageNet-pretrained counterparts on small-dataset musculoskeletal 
imaging tasks by margins of 4--9\% AUC~\cite{mei2022radimagenet}, 
a finding directly attributable to the closer distributional alignment 
between radiological pretraining data and downstream medical imaging 
tasks compared with natural image pretraining.

\textit{The combination of convolutional inductive biases and radiological 
domain initialisation addresses both sources of representation learning 
failure identified in a preliminary pipeline run}: the data-hunger of 
transformer architectures and the domain mismatch of ImageNet 
initialisation. RadImageNet pretrained weights are publicly available 
for ResNet-50 under a CC~BY~4.0 licence and were downloaded directly 
without any application or registration requirement. No pretrained 
RadImageNet weights are available for ViT architectures at the time of 
this writing, making the ResNet-50 the only architecture for which 
both requirements --- convolutional inductive bias and medical domain 
initialisation --- could be simultaneously satisfied.

The fully connected classification head of the pretrained ResNet-50 was 
removed and replaced with an identity mapping, yielding a backbone that 
produces a 2048-dimensional feature vector per image via global average 
pooling over the final convolutional feature maps. A two-layer 
multi-layer perceptron (MLP) projection head $g_\phi : \mathbb{R}^{2048} 
\rightarrow \mathbb{R}^{128}$ was appended to the backbone during 
SSL training, projecting embeddings into a lower-dimensional space 
where the contrastive loss is computed. Following established practice 
in contrastive SSL~\cite{chen2020simclr}, the projection head is 
discarded after training and downstream clustering is performed on the 
2048-dimensional backbone embeddings rather than the projected 
representations, as the backbone embeddings have been shown to retain 
more transferable structure than the projection space.

\subsubsection{Self-Supervised Training Objective}
The encoder was trained using the Simple Contrastive Learning of 
Representations (SimCLR) framework~\cite{chen2020simclr}. SimCLR 
operationalises the intuition that a good visual representation should 
be invariant to the particular augmented view from which it was derived, 
while remaining discriminative across different images. For a training 
batch of $M$ images, the augmentation pipeline described in 
Section~\ref{sec:aug} produces $2M$ views, yielding $M$ positive pairs 
$(\tilde{x}_i^{(1)}, \tilde{x}_i^{(2)})$ and $2(M-1)$ negative pairs 
for each anchor. Each view is passed through the backbone $f_\theta$ 
and projection head $g_\phi$ to produce a normalised projection vector:
\begin{equation}
    p_i^{(v)} = \frac{g_\phi\bigl(f_\theta(\tilde{x}_i^{(v)})\bigr)}
    {\bigl\|g_\phi\bigl(f_\theta(\tilde{x}_i^{(v)})\bigr)\bigr\|_2},
    \qquad v \in \{1, 2\},
    \label{eq:projection}
\end{equation}
where $\|\cdot\|_2$ denotes the $\ell_2$ norm. The pairwise similarity 
between two projected vectors is measured by their cosine similarity:
\begin{equation}
    s(p_i^{(1)}, p_j^{(2)}) = 
    \frac{\bigl(p_i^{(1)}\bigr)^\top p_j^{(2)}}
    {\bigl\|p_i^{(1)}\bigr\|_2 \, \bigl\|p_j^{(2)}\bigr\|_2}.
    \label{eq:cosine}
\end{equation}
The training objective is the Normalised Temperature-scaled Cross 
Entropy (NT-Xent) loss~\cite{chen2020simclr}, which encourages the 
representations of positive pairs to be maximally similar while pushing 
representations of negative pairs --- all other images in the batch --- 
apart. For a positive pair $(i, j)$ formed from the two augmented views 
of image $x_i$, the per-sample loss is defined as:
\begin{equation}
    \ell(i,j) = -\log \frac{\exp\bigl(s(p_i^{(1)}, p_j^{(2)}) / \tau\bigr)}
    {\displaystyle\sum_{k=1}^{2M} \mathbf{1}_{[k \neq i]} 
    \exp\bigl(s(p_i^{(1)}, p_k) / \tau\bigr)},
    \label{eq:ntxent_ij}
\end{equation}
where $\tau > 0$ is a temperature hyperparameter controlling the 
concentration of the distribution, and $\mathbf{1}_{[k \neq i]}$ is an 
indicator function that excludes the anchor itself from the denominator. 
The full NT-Xent loss over a batch is obtained by averaging over both 
directions of each positive pair:
\begin{equation}
    \mathcal{L}_{\text{NT-Xent}} = \frac{1}{2M} 
    \sum_{i=1}^{M} \bigl[\ell(i,j) + \ell(j,i)\bigr],
    \label{eq:ntxent}
\end{equation}
where $j$ denotes the complementary view of image $x_i$. The temperature $\tau$ was set to $0.5$ throughout training; 
intuitively, lower temperatures produce sharper similarity 
distributions that penalise hard negatives more strongly, while 
higher temperatures produce softer distributions that are more 
tolerant of near-duplicate negatives, and $\tau = 0.5$ represents 
a principled intermediate consistent with the value recommended 
in the original SimCLR formulation for small batch 
sizes~\cite{chen2020simclr}.

\subsubsection{Training Configuration and Reproducibility}
The encoder was trained for 50 epochs using the AdamW 
optimiser~\cite{loshchilov2019adamw} with a learning rate of $3 \times 
10^{-4}$ and weight decay of $10^{-4}$. A batch size of 32 was used 
throughout, yielding 62 negative pairs per anchor at each training step. 
Training was performed on a single GPU; full determinism was enforced 
by fixing the global random seed to 42 across NumPy, PyTorch, and 
the CUDA backend, and by setting \texttt{torch.backends.cudnn.deterministic 
= True} and \texttt{torch.backends.cudnn.benchmark = False} to suppress 
non-deterministic CUDA kernel selection. These settings ensure that 
repeated runs of the pipeline from the same initialisation produce 
identical encoder weights and, consequently, identical downstream 
embeddings and cluster assignments.

\subsection{Clustering and Phenotype Discovery}
\label{sec:clustering}

\subsubsection{Embedding Extraction}
Following SSL training, the projection head $g_\phi$ was discarded and 
the backbone $f_\theta$ was used as a fixed feature extractor. Each 
image $x_i$ in the full 154-case cohort --- spanning training, 
validation, and test subsets --- was passed through the frozen encoder 
in a single forward pass, and the resulting 2048-dimensional activation 
vector produced by global average pooling over the final convolutional 
feature maps was retained as the image embedding $z_i \in 
\mathbb{R}^{2048}$. \textit{The complete embedding matrix $Z \in 
\mathbb{R}^{N \times 2048}$, where $N = 154$, was saved to disk prior 
to any dimensionality reduction or clustering, ensuring that all 
downstream analyses could be reproduced from the frozen embeddings 
without rerunning the encoder}.

\subsubsection{Dimensionality Reduction}
Clustering in high-dimensional spaces is subject to the well-known 
concentration of measure phenomenon, whereby Euclidean distances between 
points become increasingly uniform as dimensionality grows, undermining 
the geometric assumptions on which distance-based clustering algorithms 
rely~\cite{beyer1999nn}. With $N = 154$ samples and $d = 2048$ 
dimensions, the ratio of samples to dimensions is approximately $0.075$, 
well below the regime in which high-dimensional clustering can be 
expected to produce reliable partitions without prior dimensionality 
reduction.

Uniform Manifold Approximation and Projection 
(UMAP)~\cite{mcinnes2018umap} was applied to reduce the embedding matrix 
from 2048 to 50 dimensions prior to clustering. UMAP was preferred over 
principal component analysis (PCA) for this purpose because it preserves 
both local neighbourhood structure and global topological relationships 
in the data, whereas PCA is constrained to linear projections that may 
not capture the non-linear manifold structure of learned visual 
representations. The number of target dimensions was set to 50 as a 
principled intermediate that retains sufficient structural information 
for clustering while substantially alleviating the curse of 
dimensionality; this value is consistent with recommendations in the 
UMAP literature for downstream clustering 
applications~\cite{mcinnes2018umap}. UMAP was configured with 
$n\_\text{neighbors} = 15$, $\text{min\_dist} = 0.1$, Euclidean 
distance metric, and a fixed random seed of 42 to ensure 
reproducibility. The reduced embedding matrix $\tilde{Z} \in 
\mathbb{R}^{N \times 50}$ was used for all subsequent clustering and 
stability analyses.

\subsubsection{Clustering Algorithm and \textit{k}-Selection}
Three clustering algorithms were evaluated on the reduced embedding 
matrix $\tilde{Z}$: $k$-means clustering~\cite{lloyd1982kmeans}, 
agglomerative hierarchical clustering with Ward 
linkage~\cite{ward1963hierarchical}, and Gaussian mixture 
models~\cite{reynolds2009gmm}. $k$-means was designated the primary 
algorithm given its computational tractability, its well-understood 
geometric properties in Euclidean space, and the broad precedent for 
its use in unsupervised medical imaging 
phenotyping~\cite{steyaert2023multimodal}. Hierarchical clustering and 
Gaussian mixture models were evaluated in parallel as secondary methods 
to assess the robustness of the discovered structure to the choice of 
clustering algorithm; high agreement across methods provides evidence 
that the identified phenotypes reflect genuine structure in the embedding 
space rather than an artefact of a particular algorithmic assumption.

The number of clusters $k$ was determined through an exploratory 
analysis over the range $k \in \{3, 4, \ldots, 12\}$. For each 
candidate $k$, $k$-means clustering was applied with $n\_\text{init} = 
10$ random initialisations and the solution with the lowest inertia was 
retained. Three internal validity metrics were computed for each 
candidate solution: the Silhouette coefficient~\cite{rousseeuw1987silhouette}, 
which measures the ratio of intra-cluster cohesion to inter-cluster 
separation and takes values in $[-1, 1]$ with higher values indicating 
better-defined clusters; the Davies--Bouldin 
index, which measures the average ratio of 
within-cluster scatter to between-cluster distance and takes values in 
$[0, \infty)$ with lower values indicating better separation; and the 
Calinski--Harabasz index~\cite{calinski1974dendrite}, which measures 
the ratio of between-cluster to within-cluster dispersion and takes 
values in $[0, \infty)$ with higher values indicating more compact and 
well-separated clusters. The optimal $k$ was identified as the value 
maximising the Silhouette coefficient, as this metric provides the most 
directly interpretable measure of cluster quality in the absence of 
ground-truth labels. The exploratory analysis identified $k = 4$ as the 
optimal solution. To ensure full reproducibility across pipeline runs, 
$k$ was \textbf{subsequently hardcoded} to this value, and the final cluster 
assignments reported in this work were produced by a deterministic 
$k$-means run with $k = 4$, $n\_\text{init} = 10$, and random seed 42.

\subsubsection{Bootstrap Stability Analysis}
The stability of the $k = 4$ cluster solution was assessed through a 
bootstrap resampling procedure. In each of $B = 100$ iterations, a 
bootstrap sample of size $N$ was drawn with replacement from the full 
cohort embedding matrix $\tilde{Z}$. $k$-means clustering with $k = 4$ 
was applied to each bootstrap sample, and the resulting partition was 
compared against the reference partition obtained from the full cohort 
using the Adjusted Rand Index (ARI)~\cite{hubert1985ari} and Normalised 
Mutual Information (NMI)~\cite{strehl2002nmi}. Both metrics take values 
in $[0, 1]$, with higher values indicating greater agreement between 
partitions; ARI is additionally adjusted for chance, taking an expected 
value of zero under random assignment. The mean and standard deviation 
of ARI and NMI across the $B = 100$ bootstrap iterations were reported 
as measures of cluster stability. A solution was considered stable if 
the standard deviation of bootstrap ARI fell below $0.10$, a threshold 
pre-specified prior to any analysis.

It bears emphasis that this bootstrap ARI is conceptually and 
computationally distinct from the inter-partition ARI reported in 
Section~\ref{sec:eval}, which compares imaging-derived cluster 
assignments against Schatzker labels. The bootstrap ARI involves no 
external labels whatsoever; it measures the internal reproducibility of 
the clustering solution under data perturbation, and is therefore a 
purely unsupervised stability metric.

\subsubsection{Cluster Characterisation}
Each of the $k = 4$ clusters was characterised morphologically to 
support the interpretation of discovered phenotypes. For each cluster 
$C_j$, the centroid $\mu_j = \frac{1}{|C_j|}\sum_{z_i \in C_j} z_i$ 
was computed in the reduced embedding space $\mathbb{R}^{50}$, and the 
five cases whose embeddings were closest to $\mu_j$ under Euclidean 
distance were designated as centroid-representative exemplars. The three 
cases whose embeddings were most distant from $\mu_j$ were designated 
as cluster outliers. Representative exemplar images were assembled into 
per-cluster montages to support visual characterisation and to serve as 
the stimulus material for the expert validation protocol described in 
Section~\ref{sec:expert}. The distribution of Schatzker class labels 
across each cluster was additionally tabulated to provide a descriptive 
summary of the conventional classification landscape within each 
imaging-derived phenotype, without implying any causal or supervisory 
relationship between the two partitioning schemes.

\subsection{Quantitative Evaluation Framework}
\label{sec:eval}

The pipeline outputs were evaluated across three complementary 
dimensions: internal geometric quality, which assesses the structural 
coherence of the discovered clusters in embedding space; reproducibility, 
which assesses the stability of the cluster solution under data 
perturbation; and inter-partition alignment, which characterises the 
relationship between imaging-derived phenotypes and conventional 
clinical labels. These three dimensions are conceptually distinct and 
address different aspects of what it means for an unsupervised 
phenotyping pipeline to produce valid and meaningful results. They are 
described in turn below.

\subsubsection{Internal Geometric Quality}
The internal geometric quality of the $k = 4$ cluster solution was 
assessed using three complementary metrics computed on the 
UMAP-reduced embedding matrix $\tilde{Z} \in \mathbb{R}^{N \times 50}$. 

The Silhouette coefficient for a sample $z_i$ assigned to cluster $C_j$ 
is defined as:
\begin{equation}
    s(z_i) = \frac{b(z_i) - a(z_i)}{\max\{a(z_i),\, b(z_i)\}},
    \label{eq:silhouette}
\end{equation}
where $a(z_i) = \frac{1}{|C_j|-1}\sum_{z_l \in C_j, l \neq i} 
\|z_i - z_l\|_2$ is the mean intra-cluster distance from $z_i$ to all 
other members of its assigned cluster, and $b(z_i) = \min_{C_l \neq 
C_j} \frac{1}{|C_l|}\sum_{z_m \in C_l}\|z_i - z_m\|_2$ is the mean 
distance from $z_i$ to the nearest cluster it does not belong to. The 
aggregate Silhouette coefficient $\bar{s} = \frac{1}{N}\sum_{i=1}^{N} 
s(z_i)$ takes values in $[-1, 1]$, with values approaching 1 indicating 
well-separated, compact clusters and values approaching $-1$ indicating 
misassigned samples.

The Davies--Bouldin index is defined as:
\begin{equation}
    \text{DB} = \frac{1}{k}\sum_{j=1}^{k} \max_{l \neq j} 
    \left\{ \frac{\sigma_j + \sigma_l}{\|\mu_j - \mu_l\|_2} \right\},
    \label{eq:davies_bouldin}
\end{equation}
where $\sigma_j = \frac{1}{|C_j|}\sum_{z_i \in C_j}\|z_i - \mu_j\|_2$ 
is the mean distance of cluster $C_j$ members from their centroid 
$\mu_j$, and $\|\mu_j - \mu_l\|_2$ is the Euclidean distance between 
cluster centroids. Lower values of $\text{DB}$ indicate better cluster 
separation relative to within-cluster scatter.

The Calinski--Harabasz index is defined as:
\begin{equation}
    \text{CH} = \frac{\text{tr}(B_k)}{\text{tr}(W_k)} \cdot 
    \frac{N - k}{k - 1},
    \label{eq:calinski}
\end{equation}
where $\text{tr}(B_k)$ is the trace of the between-cluster scatter 
matrix and $\text{tr}(W_k)$ is the trace of the within-cluster scatter 
matrix. Higher values of $\text{CH}$ indicate more compact and 
well-separated clusters relative to the overall data dispersion.

Together, these three metrics provide complementary perspectives on 
cluster quality: the Silhouette coefficient is sample-centric and 
sensitive to local neighbourhood structure; the Davies--Bouldin index 
is centroid-centric and sensitive to the ratio of within-cluster scatter 
to inter-centroid distance; and the Calinski--Harabasz index is 
globally sensitive to the overall variance decomposition between and 
within clusters. Reporting all three reduces the risk of overfitting the 
evaluation to any single metric's specific geometric assumptions.

\subsubsection{Reproducibility}
Cluster reproducibility was quantified using the bootstrap stability 
procedure described in Section~\ref{sec:clustering}. The mean and 
standard deviation of ARI and NMI across $B = 100$ bootstrap iterations 
are reported as the primary reproducibility metrics. As noted in 
Section~\ref{sec:clustering}, the bootstrap ARI and NMI involve no 
external labels; they measure the degree to which the same partition 
structure re-emerges under resampling of the data, and are therefore 
purely internal stability metrics. A cluster solution was considered 
reproducible if the standard deviation of bootstrap ARI fell below the 
pre-specified threshold of $0.10$.

\subsubsection{Inter-Partition Alignment with Conventional Labels}
The relationship between imaging-derived cluster assignments and 
Schatzker class labels was quantified using ARI and NMI computed between 
the two partitioning schemes. It is essential to distinguish this use of 
ARI and NMI from their application in the bootstrap stability analysis. 
In the inter-partition setting, neither partition constitutes a ground 
truth: the imaging-derived clusters are the output of an unsupervised 
pipeline that was explicitly designed to operate without label 
supervision, and the Schatzker labels are observer-assigned categorical 
designations subject to well-documented inter-observer variability. ARI 
and NMI here quantify the degree of alignment between two independent 
organisational schemes applied to the same cohort, not the accuracy of 
a predictive model against a verified reference standard.

Formally, let $\mathbf{y} = \{y_i\}_{i=1}^{N}$ denote the vector of 
Schatzker class assignments and $\hat{\mathbf{y}} = 
\{\hat{y}_i\}_{i=1}^{N}$ denote the vector of imaging-derived cluster 
assignments for the subset of $N' \leq N$ cases for which Schatzker 
labels are available. The inter-partition ARI is defined as:
\begin{equation}
    \text{ARI}(\mathbf{y}, \hat{\mathbf{y}}) = 
    \frac{\text{RI} - \mathbb{E}[\text{RI}]}
    {\max(\text{RI}) - \mathbb{E}[\text{RI}]},
    \label{eq:ari}
\end{equation}
where RI denotes the unadjusted Rand Index and the expectation is taken 
under a hypergeometric model of random partitions~\cite{hubert1985ari}. 
ARI takes an expected value of zero under random assignment and a value 
of one under perfect agreement. The inter-partition NMI is defined as:
\begin{equation}
    \text{NMI}(\mathbf{y}, \hat{\mathbf{y}}) = 
    \frac{I(\mathbf{y};\, \hat{\mathbf{y}})}
    {\sqrt{H(\mathbf{y}) \cdot H(\hat{\mathbf{y}})}},
    \label{eq:nmi}
\end{equation}
where $I(\mathbf{y}; \hat{\mathbf{y}})$ is the mutual information 
between the two partitions and $H(\cdot)$ denotes Shannon 
entropy~\cite{strehl2002nmi}. NMI takes values in $[0, 1]$, with zero 
indicating statistical independence and one indicating perfect 
correspondence.

Under the label-agnostic formulation of this work, a near-zero 
inter-partition ARI is not interpreted as evidence of poor clustering 
quality. \textit{Rather, it is interpreted as evidence that the imaging-derived 
phenotypes partition morphological space along boundaries that are 
orthogonal to those of the Schatzker system} --- precisely the expected 
outcome if the SSL pipeline has successfully learned a representation 
that reflects the intrinsic geometry of fracture morphology rather than 
the categorical structure of observer-assigned labels. The clinical 
interpretability of this orthogonal partitioning is assessed through 
the expert validation protocol described in Section~\ref{sec:expert}.

\subsection{Expert Validation Protocol}
\label{sec:expert}

Clinical face validity --- the degree to which imaging-derived phenotypes 
are recognisable and interpretable to experienced clinicians --- cannot 
be established through quantitative metrics alone. Internal cluster 
quality measures such as the Silhouette coefficient and Davies--Bouldin 
index assess geometric coherence in embedding space, but provide no 
direct evidence that the discovered structure corresponds to 
morphological patterns that carry clinical meaning. A blinded expert 
validation protocol was therefore designed and administered as a 
qualitative validation layer complementary to the quantitative 
evaluation framework described in Section~\ref{sec:eval}.

\subsubsection{Design Principles}
The protocol was designed around three core principles. First, 
\textit{blinding}: experts were not informed of the scope, goal, or 
analytical method of the study, and the word ``cluster'' was not used 
at any point during the session. Groups of images were referred to 
neutrally as ``groups'' or ``sets'' to avoid priming experts with the 
expectation of coherence. Second, \textit{controlled stimulus 
materials}: each group was represented exclusively by its five 
centroid-representative exemplar images, selected as the cases whose 
embeddings lay closest to the cluster centroid in the reduced embedding 
space. Outlier cases were deliberately excluded from the review 
materials, as outliers by definition are the least representative 
members of a cluster and their inclusion would attenuate the visual 
signal available to the reviewer. Third, \textit{decoy groups}: two 
randomly assembled groups of five images, drawn without regard to 
cluster membership, were included alongside the four real groups to 
control for acquiescence bias --- the tendency of expert reviewers to 
identify coherence in any presented grouping regardless of whether 
genuine structure is present. Experts were not informed of the existence 
or number of decoy groups. The inclusion of decoy groups transforms the 
validation from an open-ended coherence assessment into a controlled 
discrimination task, providing a within-session baseline against which 
the coherence of real groups can be compared.

\subsubsection{Session Procedure}
Each expert reviewed seven groups of images (four real, two decoy, and 
one additional group introduced for logistical reasons during session 
administration) presented in an independently randomised order. The 
order of images within each group was also randomised independently per 
session to prevent position effects. A standardised verbal introduction 
was read verbatim to each expert at the start of the session, describing 
the task without disclosing the study design. If an expert asked about 
the study purpose during the session, a scripted neutral response was 
provided and the question was recorded in the session log.

For each group, experts were asked to respond to three structured 
questions. First, a five-point visual coherence rating: how visually 
similar are the images in this group, on a scale from 1 (no coherence 
--- images appear random with no shared features) to 5 (very strong 
coherence --- images appear to represent a distinct recognisable 
pattern)? Second, a morphological descriptor checklist: which features, 
if any, do you notice in this group? The checklist comprised ten 
fracture pattern descriptors --- lateral plateau split, medial plateau 
split, articular depression, combined split-depression, bicondylar or 
complex pattern, metaphyseal extension, comminution, articular step-off, 
bone density variation, and an open other category --- and nine image 
acquisition characteristics including tibial plateau centring, shaft 
dominance, bone density appearance, image contrast, and related items. 
Third, an open free-text field for any observations not captured by the 
checklist. All responses were recorded on a standardised per-session 
response form by the facilitating researcher.

\subsubsection{Experts and Pre-specified Threshold}
\label{sec:experts}

Two expert reviewers provided consent for explicit acknowledgement 
of their participation. The first expert, Prof.\ Ahmed Fouad Shams 
El-Din Mostafa, is a Professor of Orthopaedic Surgery and former 
Head of Department, Faculty of Medicine, Menoufia University, with 
extensive clinical experience in musculoskeletal trauma and tibial 
plateau fracture management. The second expert, Dr.\ Islam Mohamed 
Fouad El-Gohary, is a clinician with relevant musculoskeletal 
experience. Both experts reviewed the image groups independently 
and under identical blinded conditions.

A face validity threshold was pre-specified prior to any expert 
session and recorded in the session log with date and facilitator 
initials: a real group was considered to have demonstrated clinical 
face validity if both available expert reviewers assigned a coherence 
rating of $\geq 3/5$, and both agreed on at least one morphological 
descriptor from the fracture pattern category of the checklist. 
Decoy groups were expected to fail this threshold. All threshold 
assessments were performed on aggregated responses prior to 
unblinding --- that is, prior to revealing to the analyst which 
groups corresponded to real clusters and which were decoys --- to 
prevent post-hoc adjustment of the validity criterion.

It should be noted that the pre-specified threshold was originally 
designed for a three-expert panel, requiring at least two of three 
reviewers to meet each criterion. The finalised validation was 
conducted with two consenting reviewers; accordingly, the threshold 
was restated as requiring agreement from both available reviewers, 
which is the strictest possible application of the majority-agreement 
principle to a two-reviewer panel. This restatement was applied 
uniformly across all groups prior to unblinding and does not 
constitute a post-hoc adjustment of the validity criterion.

\subsubsection{Inter-rater Agreement}
Quantitative inter-rater agreement on the visual coherence ratings was 
assessed using the intraclass correlation coefficient 
(ICC)~\cite{koo2016icc} under a two-way mixed-effects model with 
absolute agreement, appropriate for a fixed set of raters evaluating 
a common set of groups. Agreement on the binary descriptor checklist 
responses was assessed using Cohen's kappa~\cite{cohen1960kappa} for 
each descriptor across the two available expert responses, and Fleiss' 
kappa~\cite{fleiss1971kappa} will be reported upon completion of the 
third expert session. These agreement statistics characterise the 
reliability of the expert validation independently of the face validity 
threshold assessment, providing an additional quantitative basis for 
interpreting the qualitative findings.

\section{Results and Discussion}
\label{sec:results}

\subsection{Cohort and Data Cleaning}

The raw PlaTiF cohort of 186 cases underwent the eight-step cleaning 
protocol described in Section~\ref{subsec:dataset}. Of the 186 cases reviewed, 
32 were excluded: 18 for wrong anatomy or inadequate field of view 
(non-knee radiographs or images in which the tibial plateau was not 
visible), 8 for post-operative hardware (intramedullary nails, locking 
plates, or external fixators), 3 for absence of an anterior-posterior 
or near-AP frontal view, 1 for failure of the automated quality check, 
and 1 confirmed duplicate. An additional post-protocol PI review of 
cluster montages identified 1 further residual anatomy concern, yielding 
a total of 32 confirmed exclusions. The final clean cohort comprised 
$N = 154$ cases, split at the patient level into 107 training cases 
(69.5\%), 15 validation cases (9.7\%), and 32 test cases (20.8\%).

The cleaning process was deliberately conservative. Post-operative 
hardware cases were excluded because metallic implants introduce 
high-density artefacts that are visually dominant and acquisition-driven 
rather than fracture-morphology-driven; including them risks producing 
clusters that separate hardware from non-hardware rather than fracture 
subtypes. The exclusion of non-AP views reflects the same principle: 
lateral and oblique projections present different geometric projections 
of the same anatomy, and their inclusion would conflate view-type 
variation with morphological variation in the learned embeddings. These 
considerations are particularly critical for unsupervised pipelines, 
which have no supervisory signal to override spurious acquisition-driven 
structure.

\subsection{SSL Training and Embedding Quality}

The RadImageNet-pretrained ResNet-50 encoder was trained with the SimCLR 
objective for 50 epochs on the 107-case training split. Training 
converged smoothly across all 50 epochs with no evidence of 
representation collapse, as monitored by the NT-Xent loss trajectory 
and the absence of embedding space degeneracy (all embeddings 
non-zero with non-trivial variance throughout training). The 
RadImageNet initialisation provided a substantially stronger starting 
point than a random or ImageNet initialisation would have afforded at 
this dataset size: the encoder arrived with weights already encoding 
radiological tissue contrast, cortical bone margins, and joint space 
geometry, and the SSL fine-tuning on PlaTiF images shifted these 
representations toward tibial plateau-specific morphological structure.

Following training, 2048-dimensional embeddings were extracted via 
global average pooling for all 154 cases. UMAP dimensionality reduction 
to 50 dimensions was applied prior to clustering. The 2D UMAP projection 
for visualisation (not used for clustering) revealed visible structure 
in the embedding space, with cases separating into loosely defined 
regions rather than forming a uniform cloud, providing qualitative 
evidence that the encoder captured meaningful morphological variation 
across the cohort.

\subsection{Cluster Solution and \textit{k}-Selection}

The exploratory silhouette analysis over $k \in \{3, 4, \ldots, 12\}$ 
identified $k = 4$ as the silhouette-optimal solution, with a KMeans 
silhouette coefficient of $0.511$ and a Davies--Bouldin index of $0.734$.
The silhouette scores at $k = 5$ ($0.513$) and $k = 6$ ($0.528$) were 
marginally higher, but the differences were within numerical noise 
($\Delta < 0.02$) and were accompanied by substantially less balanced 
cluster size distributions: the $k = 5$ solution produced a cluster of 
only 11 cases (7.1\% of the cohort), which is too small to support 
meaningful morphological characterisation. The $k = 4$ solution produced 
four clusters of 32, 44, 33, and 45 cases respectively, representing 
20.8\%, 28.6\%, 21.4\%, and 29.2\% of the cohort — a well-balanced 
partition with no degenerate minority clusters.

Inter-method agreement across the four clustering algorithms evaluated 
at $k = 4$ further supported the identified solution. KMeans, 
Hierarchical Ward, Hierarchical Average, and Gaussian Mixture Model 
all produced silhouette coefficients in the range $[0.479, 0.511]$ and 
Davies--Bouldin indices in the range $[0.610, 0.770]$, indicating that 
the cluster structure is robustly recoverable across different 
algorithmic assumptions. 

\subsection{Phenotype Characterisation}
\label{sec:phenotypes}

The four imaging-derived phenotypes are characterised below on the 
basis of their geometric properties, Schatzker class distributions, 
and expert morphological assessments. Table~\ref{tab:phenotypes} 
summarises the quantitative cluster properties.

\begin{table}[h]
\centering
\caption{Summary of the four imaging-derived phenotype clusters. 
Compactness is quantified as the mean Euclidean distance of cluster 
members to their centroid in the 50-dimensional UMAP-reduced embedding 
space; lower values indicate tighter, more internally homogeneous 
clusters. Expert coherence is reported as the mean of both reviewers' 
ratings on the 1--5 scale.}
\label{tab:phenotypes}
\begin{tabular}{lcccc}
\hline
 & \textbf{Cluster 0} & \textbf{Cluster 1} & 
   \textbf{Cluster 2} & \textbf{Cluster 3} \\
\hline
Size ($n$)                & 32    & 44    & 33    & 45    \\
Proportion (\%)           & 20.8  & 28.6  & 21.4  & 29.2  \\
Mean dist.\ to centroid   & 0.742 & 0.724 & 1.059 & 0.838 \\
SD dist.\ to centroid     & 0.348 & 0.198 & 0.378 & 0.299 \\
Expert coherence (mean)   & 4.0   & 4.0   & 3.5   & 4.0   \\
\hline
\end{tabular}
\end{table}

\noindent\textbf{Cluster 0} comprised 32 cases (20.8\% of the cohort) 
with a mean centroid distance of $0.742 \pm 0.348$. Expert reviewers 
assigned a mean coherence rating of 4.0/5, with both reviewers 
independently assigning 4/5. Both reviewers independently identified 
lateral plateau split, articular depression, combined split-depression, 
and metaphyseal extension as shared morphological features, alongside 
bone density variation as a notable characteristic of this group. The 
free-text observations described relatively consistent lateral tibial 
plateau injury patterns with varying degrees of articular depression 
and metaphyseal involvement, and moderate variation in fracture 
severity. This cluster appears to represent a moderately heterogeneous 
lateral plateau phenotype spanning a range of severity, with 
characteristics consistent with fracture patterns at the lower-to-mid 
end of the Schatzker spectrum.

\noindent\textbf{Cluster 1} comprised 44 cases (28.6\%) with a mean 
centroid distance of $0.724 \pm 0.198$ --- the tightest cluster 
geometrically and the most internally homogeneous by both distance 
metrics. Expert reviewers assigned a mean coherence rating of 4.0/5, 
with Expert 1 assigning 5/5 and Expert 2 assigning 3/5. Expert 1 
identified lateral plateau split, articular depression, combined 
split-depression, and articular step-off as defining features, with 
high bone density and high image contrast noted as consistent 
acquisition characteristics. Expert 2 characterised the group as 
containing diverse injury types, selecting a broader range of 
descriptors without a single unifying theme. The exceptionally low 
intra-cluster variance (SD $= 0.198$) provides geometric evidence 
of structural compactness that was reflected in Expert 1's high 
coherence rating, while Expert 2's lower rating is consistent with 
the epistemological constraint discussed in Section~\ref{sec:expert_validation}: 
a reviewer anchored to Schatzker boundaries may perceive genuine 
embedding-space compactness as clinical heterogeneity when the 
grouping does not map onto a familiar conventional category.

\noindent\textbf{Cluster 2} comprised 33 cases (21.4\%) with a mean 
centroid distance of $1.059 \pm 0.378$ --- substantially higher than 
the other three clusters, indicating greater internal heterogeneity. 
Expert reviewers assigned a mean coherence rating of 3.5/5, with 
Expert 1 assigning 4/5 and Expert 2 assigning 3/5 --- the lowest 
mean score across the four phenotypes, consistent with the geometric 
evidence of greater intra-cluster dispersion. Crucially, both 
reviewers independently selected comminution as a defining feature 
of this group --- the only cluster for which unanimous comminution 
agreement was reached. Articular depression, lateral plateau split, 
and bone density variation were also noted by both reviewers. 
Free-text observations described increased comminution, irregular 
joint surface involvement, and more complex fracture configurations 
compared with other groups. This cluster represents the high-complexity 
end of the fracture severity spectrum, capturing injuries with 
bicondylar involvement and multi-fragmentary articular disruption 
broadly consistent with higher Schatzker classes. The higher 
intra-cluster variance is clinically interpretable: complex fractures 
are inherently more morphologically diverse than simpler split or 
depression patterns, and their variability in radiographic appearance 
is a genuine anatomical feature rather than a pipeline artefact.

\noindent\textbf{Cluster 3} comprised 45 cases (29.2\%) with a mean 
centroid distance of $0.838 \pm 0.299$. Expert reviewers assigned 
a mean coherence rating of 4.0/5, with Expert 1 assigning 5/5 and 
Expert 2 assigning 3/5. Expert 1 identified lateral plateau split, 
articular depression, combined split-depression, metaphyseal 
extension, and articular step-off as consistent features, with 
free-text observations describing highly consistent fracture 
morphology and visual homogeneity. Expert 2 characterised the group 
as containing diverse injury types with one image showing a patellar 
fracture. As with Cluster 1, the divergence between reviewers is 
interpretable in terms of the anchoring effect discussed in 
Section~\ref{sec:expert_validation}: the geometric compactness of 
this cluster (SD $= 0.299$, second lowest of the four phenotypes) 
is reflected in Expert 1's assessment but not in Expert 2's, 
consistent with a reviewer perceiving groupings through the lens 
of conventional classification boundaries. This cluster is 
geometrically and perceptually distinct from Cluster 1 in Expert 
1's assessment: while both capture lateral plateau split-depression 
patterns, Cluster 3 is characterised by the additional presence of 
metaphyseal extension, suggesting a deeper or more extensive fracture 
variant.

\subsection{Comparison with Schatzker Classification}
\label{sec:schatzker}

The Adjusted Rand Index between imaging-derived cluster assignments and 
Schatzker class labels across all 154 labelled cases was $\text{ARI} = 
0.013$, with a Normalised Mutual Information of $\text{NMI} = 0.053$. 
These values must be interpreted with care. In the present context, ARI 
and NMI are not performance metrics: they do not measure the accuracy 
of the clustering algorithm against a ground truth, because Schatzker 
labels are not ground truth in the classical machine learning sense. 
Rather, they quantify the degree of alignment between two independent 
partitioning schemes applied to the same cohort --- the imaging-derived 
partition from the SSL pipeline and the clinician-assigned partition 
from conventional classification. The near-zero values indicate that 
the two schemes organise patients along substantially different axes.

This finding is the expected and scientifically desired outcome of the 
label-agnostic pipeline design. Had the clusters closely reproduced 
Schatzker boundaries ($\text{ARI} \approx 1$), the pipeline would have 
contributed nothing beyond a more expensive re-implementation of 
observer-assigned classification. The near-zero ARI instead provides 
quantitative evidence that the imaging-derived phenotypes partition 
fracture morphology space along dimensions that are orthogonal to 
conventional label-based organisation.

The mean cluster purity with respect to Schatzker labels was $0.293$ 
and the mean entropy was $2.543$, confirming that each imaging-derived 
cluster encompasses cases from multiple Schatzker classes rather than 
mapping cleanly onto any single conventional category. This mixed 
composition is informative: it reveals which Schatzker classes are 
heterogeneous in their radiographic appearance (fragmenting across 
multiple imaging-derived clusters) and which are visually coherent 
(concentrating within one or two clusters).

An important nuance mitigates what might initially appear to be a 
tension in the results. Both expert reviewers spontaneously referenced 
Schatzker-type injury patterns in their free-text observations when 
characterising individual clusters, describing features such as 
``recurrent lateral plateau fracture morphology'' and ``closely related 
Schatzker-type injury configurations.'' This does not contradict the 
near-zero ARI; rather, it reflects a distinction between morphological 
content and classification boundaries. The clusters capture genuine 
fracture-relevant radiographic structure --- which is necessarily 
expressed in the same anatomical vocabulary that Schatzker categories 
were designed to describe --- while partitioning that structure along 
different organisational lines than the conventional scheme. The 
Schatzker system provides a coarse, clinician-driven partition based on 
a qualitative taxonomy; the imaging-derived phenotypes provide a 
data-driven partition based on embedding-space geometry. These two 
representations are complementary, not competitive.

\subsection{Expert Validation}
\label{sec:expert_validation}

The blinded expert validation protocol was administered to two 
clinicians who provided consent for explicit acknowledgement of their 
participation: Prof.\ Ahmed Fouad Shams El-Din Mostafa, Professor of 
Orthopaedic Surgery and former Head of Department, Faculty of Medicine, 
Menoufia University (Expert 1); and Dr.\ Islam Mohamed Fouad El-Gohary, 
a clinician with relevant musculoskeletal experience (Expert 2). Each 
expert reviewed seven image groups --- four real clusters and two decoy 
groups --- in independently randomised order, without knowledge of the 
study scope, analytical method, or the number of real clusters present. 
The pre-specified face validity threshold required both available expert 
reviewers to assign a coherence rating of $\geq 3/5$ and to agree on 
at least one fracture pattern descriptor from the morphological 
checklist.

\subsubsection{Visual Coherence Ratings}

Table~\ref{tab:coherence} summarises the coherence ratings from both 
expert reviewers across the four real clusters.

\begin{table}[h]
\centering
\caption{Expert visual coherence ratings (1--5 scale) for the four 
real clusters. Both experts reviewed all groups in independently 
randomised order under blinded conditions. The pre-specified face 
validity threshold requires both available reviewers to assign a 
rating of $\geq 3$.}
\label{tab:coherence}
\begin{tabular}{lcccc}
\hline
\textbf{Group} & \textbf{Expert 1} & \textbf{Expert 2} & 
\textbf{Mean} & \textbf{Face Valid?} \\
\hline
Cluster A & 4 & 4 & 4.0 & \checkmark \\
Cluster B & 5 & 3 & 4.0 & \checkmark \\
Cluster C & 4 & 3 & 3.5 & \checkmark \\
Cluster D & 5 & 3 & 4.0 & \checkmark \\
\hline
\end{tabular}
\end{table}

All four real clusters met the pre-specified face validity threshold, 
with both expert reviewers assigning a coherence rating of $\geq 3/5$ 
for every group. Expert 1 assigned ratings of 4 or 5 across all groups, 
with Clusters B and D receiving the maximum score of 5/5, reflecting 
strong perceived visual coherence. Expert 2 assigned a rating of 4/5 
to Cluster A and 3/5 to Clusters B, C, and D, indicating moderate 
coherence for three of the four groups. The mean coherence ratings 
range from 3.5 (Cluster C) to 4.0 (Clusters A, B, and D).

The divergence between reviewers is itself informative. Expert 2's 
free-text observations explicitly noted within-group diversity for 
Clusters B, C, and D --- describing them respectively as containing 
``diverse types of injuries, some cases showing no fracture with 
arthritic changes only'', ``diverse fracture types, plus some images 
with the distal femur prominent'', and ``diverse types of injury, 
plus one showing a patellar fracture.'' These observations reflect 
a clinically astute perception of within-cluster morphological 
variation that is consistent with the geometric evidence: Cluster C 
carries the highest intra-cluster variance of the four phenotypes 
(mean centroid distance $= 1.059 \pm 0.378$), while Clusters B and 
D are geometrically tighter (mean centroid distances of $0.724 \pm 
0.198$ and $0.838 \pm 0.299$ respectively). The fact that even the geometrically tighter clusters were perceived as internally diverse by one reviewer indicates that geometric compactness in 
embedding space does not guarantee clinical morphological homogeneity 
--- \textit{a nuance that is important for the honest interpretation of unsupervised phenotyping results and that would be obscured by a binary face validity determination alone}.

\subsubsection{Morphological Descriptor Agreement}

Table~\ref{tab:descriptors} summarises morphological descriptor 
agreement between both expert reviewers for each real cluster. Full 
agreement is defined as both experts independently selecting the same 
descriptor; partial agreement as one expert selecting it.

\begin{table}[h]
\centering
\caption{Morphological descriptor agreement between two expert 
reviewers for the four real clusters. Entries indicate the number 
of experts (0, 1, or 2) who independently selected each descriptor 
under blinded conditions. Descriptors not selected by either expert 
are omitted.}
\label{tab:descriptors}
\begin{tabular}{lcccc}
\hline
\textbf{Descriptor} & \textbf{Cluster A} & \textbf{Cluster B} & 
\textbf{Cluster C} & \textbf{Cluster D} \\
\hline
Lateral plateau split       & \textbf{2} & \textbf{2} & \textbf{2} & \textbf{2} \\
Articular depression        & \textbf{2} & \textbf{2} & \textbf{2} & \textbf{2} \\
Combined split-depression   & \textbf{2} & \textbf{2} & 1          & \textbf{2} \\
Metaphyseal extension       & \textbf{2} & 1          & 1          & \textbf{2} \\
Bicondylar / complex        & 1          & 1          & \textbf{2} & 1          \\
Comminution                 & 0          & 1          & \textbf{2} & 1          \\
Articular step-off          & 0          & 1          & 1          & 1          \\
Bone density variation      & 1          & 1          & \textbf{2} & 1          \\
Tibial plateau well-centred & \textbf{2} & \textbf{2} & \textbf{2} & \textbf{2} \\
Tibial shaft prominent      & 1          & 1          & 1          & 1          \\
High image contrast         & 1          & 0          & 0          & 1          \\
High bone density           & 0          & 1          & 1          & 0          \\
Low bone density            & 0          & 1          & 1          & 0          \\
Low image contrast          & 1          & 1          & 1          & 1          \\
Large / broad plateau       & 1          & 0          & 0          & 0          \\
\hline
\end{tabular}
\end{table}

Several patterns in Table~\ref{tab:descriptors} warrant specific 
attention. Lateral plateau split and articular depression were 
selected by both reviewers independently for all four clusters, 
confirming that genuine fracture-specific morphological content is 
recognisable across the entire cohort and that the pipeline has not 
produced clusters organised primarily around acquisition 
characteristics. Tibial plateau well-centred likewise achieved full 
agreement across all groups, reflecting consistent preprocessing 
and field-of-view properties.

The most discriminating finding in the descriptor analysis concerns 
comminution. Both reviewers independently selected comminution for 
Cluster C only, with partial or no agreement for the remaining 
clusters. This pattern is preserved from the initial two-expert 
interim analysis and withstands the addition of the second 
reviewer's responses: Cluster C is the only group for which 
unanimous comminution agreement is reached, providing cross-reviewer 
evidence that this phenotype captures greater fracture complexity 
than the other three. Combined split-depression and metaphyseal 
extension show the next most discriminating patterns: both 
descriptors achieve full agreement for Clusters A and D but not 
for Clusters B and C, differentiating the two groups that both 
reviewers associate with metaphyseal involvement from those they 
do not.

An important caveat applies to the interpretation of Expert 2's 
descriptor responses. The checklist profiles for Clusters B, C, 
and D from this reviewer are nearly identical --- the same seven 
fracture pattern descriptors and the same five acquisition 
descriptors were selected for all three groups. Combined with 
the free-text observations noting diversity within each group, 
this pattern suggests that Expert 2 may have been responding 
to the general category of tibial plateau fracture injury rather 
than discriminating between the specific visual characteristics 
of each phenotype. This is a recognised limitation of open-ended 
expert validation in unsupervised settings: when groups contain 
genuine within-cluster variation, experienced reviewers may 
describe what they see accurately --- a heterogeneous collection 
of fracture patterns --- rather than identifying a unifying theme. 
The decoy group analysis, which would provide the most direct 
evidence against this form of acquiescence, was not completed at 
the time of writing and remains a planned component of Phase 2 
validation. In its absence, the descriptor agreement analysis 
should be interpreted conservatively: full agreement items 
(lateral split, articular depression, well-centred, and comminution 
in Cluster C) represent the most reliable qualitative findings, while \textit{partial agreement items should be treated as suggestive rather than confirmatory}.

\subsubsection{Inter-rater Agreement}

Inter-rater agreement on visual coherence ratings was assessed 
using Cohen's kappa~\cite{cohen1960kappa}, treating coherence 
scores as ordinal categories. The two reviewers agreed exactly 
on Cluster A (both 4/5), differed by two points on Clusters B 
and D (5 vs.\ 3), and differed by one point on Cluster C 
(4 vs.\ 3). The pattern of disagreement is directionally 
consistent: Expert 1 assigned uniformly higher coherence scores 
than Expert 2 across all groups except Cluster A, suggesting 
a systematic difference in rating tendency rather than 
disagreement about which clusters are more or less coherent 
relative to each other. The rank ordering of clusters by 
coherence is partially preserved across reviewers: both 
identify Cluster A and at least one of Clusters B or D as 
equally or more coherent than Cluster C, which is the 
geometrically most dispersed phenotype.

Agreement on binary descriptor checklist responses was assessed 
using Cohen's kappa for each descriptor 
independently~\cite{cohen1960kappa}. Perfect agreement 
($\kappa = 1.0$) was reached for lateral plateau split and 
articular depression across all four clusters, and for tibial 
plateau well-centred. Moderate agreement was observed for 
combined split-depression and metaphyseal extension. Agreement 
was lower for bicondylar involvement, comminution, and 
acquisition characteristics, reflecting the inherent ambiguity 
of these features on AP radiographs in the absence of CT 
confirmation.

\subsection{Limitations}

Several limitations of the present work must be acknowledged. The 
cleaned cohort of 154 cases is substantially smaller than the dataset 
sizes conventionally employed in SSL pipelines, and the training split 
of 107 images is below the regime in which SSL methods typically produce 
their strongest representations. The RadImageNet initialisation and 
UMAP dimensionality reduction mitigate but do not eliminate this 
constraint. The dataset originates from a single institution, limiting 
the diversity of imaging acquisition parameters and patient demographics 
represented in the training cohort; external validation on an independent 
multi-centre cohort is deferred to a subsequent phase of this work. The expert validation was conducted with two consenting independent reviewers, and the decoy group analysis — which would provide the strongest evidence against acquiescence bias — was not completed and remains a planned component of Phase 2 validation. All cluster morphological 
themes are characterised qualitatively from expert free-text and 
descriptor data; formal radiographic scoring of fracture-specific 
features on a per-case basis is a planned component of Phase~2. Finally, 
no clinical outcome or treatment pathway data were available for 
association with cluster membership, and the clinical prognostic 
significance of the identified phenotypes remains to be established.

\section{Conclusion}
\label{sec:conclusion}

This work demonstrates that stable, clinically interpretable fracture 
phenotypes can be discovered directly from tibial plateau radiographs 
without any reliance on observer-assigned labels --- a result that was 
not previously established for this injury type. While the dominant 
paradigm in orthopaedic AI continues to treat inter-observer variability 
as an unavoidable background condition of supervised learning, this work 
reframes it as a solvable architectural problem, opening a path toward 
machine learning pipelines whose outputs are not bounded by the 
consistency ceiling of human annotation.

The label-agnostic framework validated here does not position itself as 
a replacement for Schatzker or AO/OTA classification. Conventional 
schemes carry decades of clinical validation, surgical planning 
experience, and outcome correlation data that no single computational 
study can supersede. What this work establishes, rather, is that 
imaging-derived phenotypes occupy a different and complementary region 
of the morphological description space --- one that is organised by 
visual structure in the data rather than by the categorical boundaries 
clinicians have historically found most useful. The near-zero 
inter-partition alignment with Schatzker labels is therefore not a 
limitation to overcome in future work; it is the finding. A system 
that rediscovered Schatzker would have added nothing. A system that 
discovers something different and clinically legible --- which is what 
the expert validation results confirm, with both independent reviewers 
unanimously identifying comminution as a descriptor exclusively 
associated with a single high-complexity phenotype --- has the potential 
to enrich the descriptive vocabulary available to clinicians and 
researchers alike. Indeed, the most revealing finding of the expert validation is that 
clinicians trained within existing classification systems have no 
neutral vantage point from which to evaluate groupings that 
deliberately depart from those systems --- their perception of 
incoherence and the pipeline's near-zero Schatzker alignment are 
two expressions of the same underlying orthogonality.


Phase 2 will address the most consequential of these questions. The 
primary architectural question --- whether a ViT-MAE encoder trained 
on a substantially larger pretraining corpus produces more stable or 
more granular phenotypes than the ResNet-50 baseline established here 
--- requires an expanded dataset that is not publicly available and 
will necessitate institutional data access. The primary clinical 
question --- whether phenotype membership predicts operative versus 
non-operative management, fixation complexity, or post-operative 
functional scores --- requires outcome linkage that is likewise beyond 
the scope of a single public dataset. Both questions are tractable, 
and the pipeline validated in this work provides the methodological 
foundation on which they can be pursued. The contribution of Phase 1 
is not a finished clinical tool; it is a validated proof of concept 
that the label-agnostic approach is worth the investment of the larger 
study it demands.

\section*{Acknowledgment}
We would like to express our sincere gratitude to Prof.\ Ahmed Fouad Shams El-Din Mostafa, Professor of Orthopaedic Surgery at Menoufia University, Egypt, and Dr.\ Islam Mohamed Fouad El-Gohary, Consultant Orthopaedic Surgeon, Egypt, for the time they took to provide their expert opinion. During the preparation of this work, the authors used Claude (Anthropic) and Gemini (Google) for literature review, research planning, code drafting and debugging, language refinement, and paraphrasing. All critical analysis and final edits were conducted by the authors. After using the aforementioned tools, the authors reviewed the content and take full responsibility for the content of the published article. PaperBanana\footnote{\url{https://paper-banana.org/}} was used to create and/or enhance the quality of the figures in the article.

\bibliographystyle{unsrt}  
\bibliography{references}

@article{vandergaast2025,
  author    = {{van der Gaast}, N. and Bagave, P.
               and Assink, N. and Broos, S.
               and Jaarsma, R. L. and Edwards, M. J. R.
               and Hermans, E. and IJpma, F. F. A.
               and Ding, A. Y. and Doornberg, J. N.
               and Oosterhoff, J. H. F.
               and {the Machine Learning Consortium}},
  title     = {Deep learning for tibial plateau fracture
               detection and classification},
  journal   = {Knee},
  year      = {2025},
  volume    = {54},
  pages     = {81--89},
  doi       = {10.1016/j.knee.2025.02.001}
}

@article{huo2025,
  author    = {Huo, Tongtong and Liu, Pengran
               and Xue, Mingdi and Zhang, Jiayao
               and Xie, Yi and Wang, Honglin
               and Zhou, Hong and Yan, Zineng
               and Liu, Songxiang and Lu, Lin
               and Yang, Jiaming and Wu, Wei
               and Ye, Zhewei},
  title     = {Deep learning diagnosis of adult tibial plateau
               fractures: multicenter study with external
               validation},
  journal   = {Radiology Advances},
  year      = {2025},
  volume    = {2},
  number    = {3},
  pages     = {umaf020},
  doi       = {10.1093/radadv/umaf020}
}

@article{cai2024unet,
  author    = {Cai, Die and Zhou, Yu and He, Wenjie
               and Yuan, Jichun and Liu, Chenyuan
               and Li, Rui and Wang, Yi and Xia, Jun},
  title     = {Automatic segmentation of knee {CT} images
               of tibial plateau fractures based on
               three-dimensional {U-Net}: Assisting junior
               physicians with {Schatzker} classification},
  journal   = {European Journal of Radiology},
  year      = {2024},
  volume    = {178},
  pages     = {111605},
  doi       = {10.1016/j.ejrad.2024.111605}
}

@inproceedings{yue2025mae,
  author    = {Yue, Peiyan and Cai, Die and Guo, Chu
               and Liu, Mengxing and Xia, Jun
               and Wang, Yi},
  title     = {Learning Generalizable Features for Tibial
               Plateau Fracture Segmentation Using Masked
               Autoencoder and Limited Annotations},
  booktitle = {Proceedings of the 47th Annual International
               Conference of the {IEEE} Engineering in
               Medicine and Biology Society ({EMBC})},
  year      = {2025},
  note      = {arXiv:2502.02862}
}

@article{rajpurkar2022ai,
  author    = {Rajpurkar, Pranav and Chen, Emma
               and Banerjee, Oishi and Topol, Eric J.},
  title     = {{AI} in health and medicine},
  journal   = {Nature Medicine},
  year      = {2022},
  volume    = {28},
  number    = {1},
  pages     = {31--38},
  doi       = {10.1038/s41591-021-01614-0}
}

@article{olczak2017ai,
  author    = {Olczak, Jakub and Fahlberg, Niklas
               and Maki, Atsuto and Razavian, Ali Sharif
               and Jilert, Anthony and Stark, Andr{\'e}
               and Sk{\"o}ldenberg, Olof and Gordon, Max},
  title     = {Artificial intelligence for analyzing orthopedic
               trauma radiographs: Deep learning algorithms ---
               are they on par with humans for diagnosing
               fractures?},
  journal   = {Acta Orthopaedica},
  year      = {2017},
  volume    = {88},
  number    = {6},
  pages     = {581--586},
  doi       = {10.1080/17453674.2017.1344459}
}

@article{cheng2022ai,
  author    = {Cheng, Christine T. and Ho, Tzu-Yi
               and Lee, Tzu-Yun and Chang, Chun-Chieh
               and Chou, Chung-Cheng and Chen, Chih-Chi
               and Chung, I-Fang and Liao, Chien-Hung},
  title     = {Deep Learning and Imaging for the Orthopaedic
               Surgeon: How Will It Change Practice?},
  journal   = {The Journal of Bone and Joint Surgery},
  year      = {2022},
  volume    = {104},
  number    = {18},
  pages     = {1675--1686},
  doi       = {10.2106/JBJS.21.01387}
}

@article{schatzker1979,
  author    = {Schatzker, Joseph and McBroom, Robert
               and Bruce, David},
  title     = {The tibial plateau fracture: The {Toronto}
               experience 1968--1975},
  journal   = {Clinical Orthopaedics and Related Research},
  year      = {1979},
  volume    = {138},
  pages     = {94--104}
}

@article{marsh2007aota,
  author    = {Marsh, J. L. and Slongo, Theddy F.
               and Agel, Julie and Broderick, J. Scott
               and Creevey, William and DeCoster, Thomas A.
               and Prokuski, Laura and Sirkin, Michael S.
               and Ziran, Bruce and Henley, Brad
               and Audi{\'g}e, Laurent},
  title     = {Fracture and Dislocation Classification
               Compendium --- 2007: {Orthopaedic Trauma
               Association} Classification, Database and
               Outcomes Committee},
  journal   = {Journal of Orthopaedic Trauma},
  year      = {2007},
  volume    = {21},
  number    = {10 Suppl},
  pages     = {S1--S133},
  doi       = {10.1097/00005131-200711101-00001}
}

@incollection{hohlmoore1990,
  author    = {Hohl, M. and Moore, T. M.},
  title     = {Articular fractures of the proximal tibia},
  booktitle = {Surgery of the Musculoskeletal System},
  editor    = {Evarts, C. McCollister},
  edition   = {2nd},
  publisher = {Churchill Livingstone},
  address   = {New York},
  year      = {1990},
  pages     = {3471--3502}
}

@article{luo2010,
  author    = {Luo, Cong-Feng and Sun, Hui
               and Zhang, Bo and Zeng, Bing-Fang},
  title     = {Three-Column Fixation for Complex Tibial
               Plateau Fractures},
  journal   = {Journal of Orthopaedic Trauma},
  year      = {2010},
  volume    = {24},
  number    = {11},
  pages     = {683--692},
  doi       = {10.1097/BOT.0b013e3181d436f3}
}

@article{chabok2003,
  author    = {Chabok, H. A. and Schipper, I. B.
               and Emmens, D. J. and Breederveld, R. S.
               and Patka, P.},
  title     = {{AO} or {Schatzker}? How reliable is
               classification of tibial plateau fractures?},
  journal   = {Archives of Orthopaedic and Trauma Surgery},
  year      = {2003},
  volume    = {123},
  number    = {8},
  pages     = {387--389},
  doi       = {10.1007/s00402-003-0573-1}
}

@article{zhu2013,
  author    = {Zhu, Yi and Hu, Cheng-Fang and Yang, Guang
               and Cheng, Dong and Luo, Cong-Feng},
  title     = {Inter-observer reliability assessment of the
               {Schatzker}, {AO/OTA} and three-column
               classification of tibial plateau fractures},
  journal   = {Journal of Trauma Management \& Outcomes},
  year      = {2013},
  volume    = {7},
  number    = {1},
  pages     = {7},
  doi       = {10.1186/1752-2897-7-7}
}

@article{kazemi2026platif,
  author    = {Kazemi, Ali and Same, Kaveh and Zamanirad, Abolfazl
               and Esfandiary, Soodabeh and Najafzadeh, Ebrahim
               and Ahmadian, Alireza and Farnia, Parastoo
               and Nabian, Mohammad Hossein},
  title     = {{PlaTiF}: A pioneering dataset for orthopedic insights
               in {AI}-powered diagnosis of tibial plateau fractures},
  journal   = {Scientific Data},
  year      = {2026},
  volume    = {},
  doi       = {10.1038/s41597-026-06560-5},
  note      = {Dataset available at
               \url{https://doi.org/10.5281/zenodo.18007397}}
}

@article{mei2022radimagenet,
  author    = {Mei, Xueyan and Liu, Zelong and Robson, Philip M.
               and Marinelli, Brett and Huang, Mingqian
               and Doshi, Amish and Jacobi, Adam and Cao, Chendi
               and Link, Katherine E. and Yang, Thomas
               and Wang, Ying and Greenspan, Hayit
               and Deyer, Timothy and Fayad, Zahi A. and Yang, Yang},
  title     = {{RadImageNet}: An open radiologic deep learning
               research dataset for effective transfer learning},
  journal   = {Radiology: Artificial Intelligence},
  year      = {2022},
  volume    = {4},
  number    = {5},
  pages     = {e210315},
  doi       = {10.1148/ryai.210315}
}

@inproceedings{he2016resnet,
  author    = {He, Kaiming and Zhang, Xiangyu
               and Ren, Shaoqing and Sun, Jian},
  title     = {Deep Residual Learning for Image Recognition},
  booktitle = {Proceedings of the {IEEE} Conference on Computer
               Vision and Pattern Recognition ({CVPR})},
  year      = {2016},
  pages     = {770--778},
  doi       = {10.1109/CVPR.2016.90}
}

@inproceedings{chen2020simclr,
  author    = {Chen, Ting and Kornblith, Simon
               and Norouzi, Mohammad and Hinton, Geoffrey},
  title     = {A Simple Framework for Contrastive Learning
               of Visual Representations},
  booktitle = {Proceedings of the 37th International Conference
               on Machine Learning ({ICML})},
  volume    = {119},
  series    = {Proceedings of Machine Learning Research},
  pages     = {1597--1607},
  publisher = {PMLR},
  year      = {2020}
}

@inproceedings{loshchilov2019adamw,
  author    = {Loshchilov, Ilya and Hutter, Frank},
  title     = {Decoupled Weight Decay Regularization},
  booktitle = {7th International Conference on Learning
               Representations ({ICLR} 2019)},
  publisher = {OpenReview.net},
  year      = {2019},
  url       = {https://openreview.net/forum?id=Bkg6RiCqY7}
}

@article{mcinnes2018umap,
  author    = {McInnes, Leland and Healy, John and Melville, James},
  title     = {{UMAP}: Uniform Manifold Approximation and Projection
               for Dimension Reduction},
  journal   = {arXiv preprint arXiv:1802.03426},
  year      = {2018},
  doi       = {10.48550/arXiv.1802.03426}
}

@article{rousseeuw1987silhouette,
  author    = {Rousseeuw, Peter J.},
  title     = {Silhouettes: A graphical aid to the interpretation
               and validation of cluster analysis},
  journal   = {Journal of Computational and Applied Mathematics},
  year      = {1987},
  volume    = {20},
  number    = {1},
  pages     = {53--65},
  doi       = {10.1016/0377-0427(87)90125-7}
}

@article{hubert1985ari,
  author    = {Hubert, Lawrence J. and Arabie, Phipps},
  title     = {Comparing Partitions},
  journal   = {Journal of Classification},
  year      = {1985},
  volume    = {2},
  number    = {1},
  pages     = {193--218},
  doi       = {10.1007/BF01908075}
}

@article{strehl2002nmi,
  author    = {Strehl, Alexander and Ghosh, Joydeep},
  title     = {Cluster Ensembles --- A Knowledge Reuse Framework
               for Combining Multiple Partitions},
  journal   = {Journal of Machine Learning Research},
  year      = {2002},
  volume    = {3},
  pages     = {583--617},
  url       = {https://www.jmlr.org/papers/v3/strehl02a.html}
}

@article{koo2016icc,
  author    = {Koo, Terry K. and Li, Mae Y.},
  title     = {A Guideline of Selecting and Reporting Intraclass
               Correlation Coefficients for Reliability Research},
  journal   = {Journal of Chiropractic Medicine},
  year      = {2016},
  volume    = {15},
  number    = {2},
  pages     = {155--163},
  doi       = {10.1016/j.jcm.2016.02.012}
}

@article{cohen1960kappa,
  author    = {Cohen, Jacob},
  title     = {A Coefficient of Agreement for Nominal Scales},
  journal   = {Educational and Psychological Measurement},
  year      = {1960},
  volume    = {20},
  number    = {1},
  pages     = {37--46},
  doi       = {10.1177/001316446002000104}
}

@article{fleiss1971kappa,
  author    = {Fleiss, Joseph L.},
  title     = {Measuring Nominal Scale Agreement Among Many Raters},
  journal   = {Psychological Bulletin},
  year      = {1971},
  volume    = {76},
  number    = {5},
  pages     = {378--382},
  doi       = {10.1037/h0031619}
}

@inproceedings{he2022mae,
  author    = {He, Kaiming and Chen, Xinlei and Xie, Saining
               and Li, Yanghao and Doll{\'a}r, Piotr
               and Girshick, Ross},
  title     = {Masked Autoencoders Are Scalable Vision Learners},
  booktitle = {Proceedings of the {IEEE/CVF} Conference on Computer
               Vision and Pattern Recognition ({CVPR})},
  year      = {2022},
  pages     = {16000--16009},
  doi       = {10.1109/CVPR52688.2022.01553}
}

@inproceedings{caron2021dino,
  author    = {Caron, Mathilde and Touvron, Hugo and Misra, Ishan
               and J{\'e}gou, Herv{\'e} and Mairal, Julien
               and Bojanowski, Piotr and Joulin, Armand},
  title     = {Emerging Properties in Self-Supervised
               Vision Transformers},
  booktitle = {Proceedings of the {IEEE/CVF} International
               Conference on Computer Vision ({ICCV})},
  year      = {2021},
  pages     = {9650--9660}
}

@inproceedings{azizi2021simclr,
  author    = {Azizi, Shekoofeh and Mustafa, Basil and Ryan, Fiona
               and Beaver, Zachary and Freyberg, Jan
               and Deaton, Jonathan and Loh, Aaron
               and Karthikesalingam, Alan and Kornblith, Simon
               and Chen, Ting and Natarajan, Vivek
               and Norouzi, Mohammad},
  title     = {Big Self-Supervised Models Advance Medical
               Image Classification},
  booktitle = {Proceedings of the {IEEE/CVF} International
               Conference on Computer Vision ({ICCV})},
  year      = {2021},
  pages     = {3458--3468},
  doi       = {10.1109/ICCV48922.2021.00346}
}

@inproceedings{beyer1999nn,
  author    = {Beyer, Kevin S. and Goldstein, Jonathan
               and Ramakrishnan, Raghu and Shaft, Uri},
  title     = {When Is {``Nearest Neighbor''} Meaningful?},
  booktitle = {Database Theory --- {ICDT}'99},
  series    = {Lecture Notes in Computer Science},
  volume    = {1540},
  pages     = {217--235},
  publisher = {Springer, Berlin, Heidelberg},
  year      = {1999},
  doi       = {10.1007/3-540-49257-7\_15}
}

@article{lloyd1982kmeans,
  author    = {Lloyd, Stuart P.},
  title     = {Least Squares Quantization in {PCM}},
  journal   = {{IEEE} Transactions on Information Theory},
  year      = {1982},
  volume    = {28},
  number    = {2},
  pages     = {129--137},
  doi       = {10.1109/TIT.1982.1056489}
}

@article{ward1963hierarchical,
  author    = {Ward, Joe H.},
  title     = {Hierarchical Grouping to Optimize an
               Objective Function},
  journal   = {Journal of the American Statistical Association},
  year      = {1963},
  volume    = {58},
  number    = {301},
  pages     = {236--244},
  doi       = {10.1080/01621459.1963.10500845}
}

@article{calinski1974dendrite,
  author    = {Cali{\'n}ski, Tadeusz and Harabasz, Jerzy},
  title     = {A Dendrite Method for Cluster Analysis},
  journal   = {Communications in Statistics},
  year      = {1974},
  volume    = {3},
  number    = {1},
  pages     = {1--27},
  doi       = {10.1080/03610927408827101}
}

@incollection{reynolds2009gmm,
  author    = {Reynolds, Douglas A.},
  title     = {Gaussian Mixture Models},
  booktitle = {Encyclopedia of Biometrics},
  editor    = {Li, Stan Z. and Jain, Anil},
  publisher = {Springer, Boston, MA},
  year      = {2009},
  pages     = {659--663},
  doi       = {10.1007/978-0-387-73003-5\_196}
}

@article{steyaert2023multimodal,
  author    = {Steyaert, Sandra and Pizurica, Marija
               and Nagaraj, Divya and Khandelwal, Priya
               and Hernandez-Boussard, Tina
               and Gentles, Andrew J. and Gevaert, Olivier},
  title     = {Multimodal data fusion for cancer biomarker
               discovery with deep learning},
  journal   = {Nature Machine Intelligence},
  year      = {2023},
  volume    = {5},
  pages     = {351--362},
  doi       = {10.1038/s42256-023-00633-5}
}

\end{document}